\documentclass[]{JHEP3}
\usepackage{epsfig}

\def\be{\begin{equation}}
\def\ee{\end{equation}}
\def\bea{\begin{array}}
\def\eea{\end{array}}
\def\beqa{\begin{eqnarray}}
\def\eeqa{\end{eqnarray}}
\def\beqas{\begin{eqnarray*}}
\def\eeqas{\end{eqnarray*}}

\def\bp{\begin{picture}}
\def\ep{\end{picture}}
\def\bc{\begin{center}}
\def\ec{\end{center}}
\def\bfig{\begin{figure}}
\def\efig{\end{figure}}

\def\bit{\begin{itemize}}
\def\eit{\end{itemize}}
\def\nn{\nonumber}
\def\f{\frac}

\def\[{\left[}
\def\]{\right]}
\def\({\left(}
\def\){\right)}

\def\..{\left.}
\def\.{\right.}
\def\tl{\tilde}
\def\ra{\rightarrow}
\def\la{\leftarrow}

\def\tm{\times}

\def\da{\dagger}

\def\la{\lambda}

\def\ep{\epsilon}

\def\Ga{\Gamma}

\def\pa{\partial}
\def\pr{\prime}
\def\eqv{\equiv}

\title{$SU(7)$ Unification of $SU(3)_C\tm SU(4)_W\tm U(1)_{B-L}$}
\author{Csaba Bal\'azs$^1$, Tianjun Li$^{2,3}$, Fei Wang$^1$ and Jin Min Yang$^2$ \\
 $^1$ School of Physics, Monash University, Melbourne Victoria 3800, Australia\\
 $^2$ Key Laboratory of Frontiers in Theoretical Physics,
      Institute of Theoretical Physics, Chinese Academy of Sciences,
      Beijing 100190, P. R. China \\
$^3$ George P. and Cynthia W. Mitchell Institute for Fundamental
Physics,
     Texas A$\&$M University, College Station, TX 77843, USA
}

\abstract{ We propose the SUSY $SU(7)$ unification of the $SU(3)_C\tm SU(4)_W\tm
  U(1)_{B-L}$ model. Such unification scenario has rich symmetry breaking chains
  in a five-dimensional orbifold. We study in detail the SUSY $SU(7)$ symmetry breaking
  into $SU(3)_C\tm SU(4)_W\tm U(1)_{B-L}$ by boundary conditions in a Randall-
  Sundrum background and its AdS/CFT interpretation.
  We find that successful gauge coupling unification can be achieved in our
  scenario. Gauge unification favors low left-right and unification scales with
  tree-level $\sin^2\theta_W=0.15$.
  We use the AdS/CFT dual of the conformal supersymmetry breaking scenario to
  break the remaining ${\cal N}=1$ supersymmetry. We employ AdS/CFT to reproduce the NSVZ
  formula and obtain the structure of the Seiberg duality in the strong coupling
  region for $\f{3}{2}N_c<N_F<3N_C$. We show that supersymmetry is indeed broken
  in the conformal supersymmetry breaking scenario with a vanishing singlet
  vacuum expectation value. }

\begin{document}
\indent
\newpage
\section{Introduction}
The standard model (SM) of electroweak interactions, based on the
spontaneously broken $SU(2)_L{\times}U(1)_Y$ gauge symmetry, has
been extremely successful in describing phenomena below the weak
scale.  However, the SM leaves some theoretical and aesthetical
questions unanswered, two of which are the origin of parity
violation and the smallness of neutrino masses.  Both of these
questions can be addressed in the left-right model based on the
$SU(2)_L\times SU(2)_R\tm U(1)_{B-L}$ gauge symmetry~\cite{mohapatra}. The
supersymmetric extension of this model~\cite{susylr} is especially
intriguing since it automatically preserves R-parity.  This can lead
to a low energy theory without baryon number violating interactions
after R-parity is spontaneously broken.  However, in such left-right
models parity invariance and the equality of the $SU(2)_L$ and
$SU(2)_R$ gauge couplings is ad hoc and has to be imposed by hand.
Only in Grand Unified Theories (GUTs)~\cite{su5,so10} can the equality of
the two $SU(2)$ gauge couplings be naturally guaranteed through
gauge coupling unification.

Novel attempts for the unification of the left-right symmetries have
been proposed in the literature, such as the $SU(3)_C\tm
SU(4)_W\tm U(1)_{B-L}$~\cite{nandi,fei1,shafi} or $SU(3)_C\tm
SU(4)_W$~\cite{fei2}. In these attempts, the equality of the
left-right gauge couplings and the parity in the left-right model are
understood by partial unification. In this work, we propose to embed the
$SU(3)_C\tm SU(4)_W\tm U(1)_{B-L}$ partially unified model
into an $SU(7)$ GUT. Unfortunately, the
doublet-triplet splitting problem exists in various GUT
models. An elegant solution to this is to invoke a higher dimensional
space-time and to break the GUT symmetry by boundary conditions
such as orbifold projection.
Orbifold GUT models for $SU(5)$ were proposed
in~\cite{Kawamura:1999nj, Kawamura:2000ev,Kawamura:2000ir} and
widely studied thereafter
in~\cite{at,Hall:2001pg,Kobakhidze:2001yk,Hebecker:2001wq,
Hebecker:2001jb,Li:2001qs, Li:2001wz,fei3,fei4}. The embedding of
the supersymmetric GUT group into the Randall-Sundrum (RS) model~\cite{rs}
with a warped extra dimension is especially interesting since it has a
four-dimensional (4D) conformal field theory (CFT)
interpretation~\cite{nomura2,nomura3}.
By assigning different symmetry breaking
boundary conditions to the two fixed point, the
five-dimensional (5D) theory is
interpreted to be the dual of a 4D technicolor-like theory or a
composite gauge symmetry model.

 It is desirable to introduce supersymmetry in warped space-time\cite{tonysusy} because we can not only stabilize the
gauge hierarchy by supersymmetry but also set the supersymmetry broken scale
by warping. It is well known that supersymmetry (SUSY) can be broken by selecting proper
boundary conditions in the high dimensional theory. For example,
5D ${\cal N}=1$ supersymmetry, which amounts to ${\cal N}=2$ supersymmetry in 4D, can be
broken to 4D ${\cal N}=1$ supersymmetry by orbifold projection. Various mechanisms
can be used to break the remaining ${\cal N}=1$ supersymmetry. One
intriguing possibility is the recently proposed conformal
supersymmetry breaking mechanism~\cite{yanagida1,yanagida2} in
vector-like gauge theories which can be embedded into a semi-direct
gauge mediation model.  Such a semi-direct gauge mediation model
can be very predictive having only one free parameter. It
is interesting to recast it in a warped extra dimension via the
Anti-de Sitter/Conformal Field Theory (AdS/CFT)
correspondence~\cite{maldacena}, and use it to break the remaining
supersymmetry.

This paper is organized as follows. In Section~\ref{sec-1}, as a warm up, we
discuss the SUSY $SU(7)$ orbifold GUT model and its symmetry breaking chains
in a flat extra dimension. In Section~\ref{sec-2}, we present the
SUSY $SU(7)$ GUT model with a warped extra dimension and its 4D CFT dual
interpretation. In Section~\ref{sec-3} we consider gauge
coupling unification in the RS background. In Section~\ref{sec-4} we
discuss the AdS/CFT dual of the semi-direct gauge mediation model in the
conformal window in vector-like gauge theories. Section~\ref{sec-5}
contains our conclusions.

\section{SUSY $SU(7)$ Unification in a Flat Extra Dimension}
\label{sec-1}
 We consider ${\cal M}_4{\tm} S^1/Z_2$, the 5D space-time
comprising of the Minkowski space ${\cal M}_4$
with coordinates $x_{\mu}$ and the orbifold $S^1/Z_2$ with
coordinate $y \eqv x_5$. The orbifold $S^1/Z_2$ is obtained from
$S^1$ by moduling the equivalent classes \beqa Z_5:~ y&\ra&-y~.
\eeqa There are two inequivalent 3-branes located at $y=0$ and $y=\pi R$,
denoted by $O$ and $O^{\pr}$, respectively.

The 5D ${\cal N}=1$ supersymmetric gauge theory has 8 real
supercharges, corresponding to ${\cal N}=2$ SUSY in 4D.
The vector multiplet contains a vector boson
$A_M$ ($M=0, 1, 2, 3, 5$), two Weyl gauginos $\lambda_{1,2}$,
and a real scalar $\sigma$. From the 4D ${\cal N}=1$ point of view,
it contains a vector multiplet $V(A_{\mu}, \lambda_1)$ and
a chiral multiplet $\Sigma((\sigma+iA_5)/\sqrt 2, \lambda_2)$ which
transform in the adjoint representation of the gauge group.
The 5D hypermultiplet contains two complex scalars
$\phi$ and $\phi^c$, a Dirac fermion $\Psi$, and can be decomposed
into two 4D chiral mupltiplets $\Phi(\phi, \psi \equiv
\Psi_R)$ and $\Phi^c(\phi^c, \psi^c \equiv \Psi_L)$, which are
conjugates of each other under gauge transformations.
The general action for the gauge fields and their couplings to the
bulk hypermultiplet $\Phi$ is~\cite{nima,nima2}
\begin{eqnarray}
S&=&\int{d^5x}\frac{1}{k g^2} {\rm
Tr}\left[\frac{1}{4}\int{d^2\theta} \left(W^\alpha W_\alpha+{\rm H.
C.}\right) \right.\nonumber\\&&\left.
+\int{d^4\theta}\left((\sqrt{2}\partial_5+ {\bar \Sigma })
e^{-V}(-\sqrt{2}\partial_5+\Sigma )e^V+
\partial_5 e^{-V}\partial_5 e^V\right)\right]
\nonumber\\&& +\int{d^5x} \left[ \int{d^4\theta} \left( {\Phi}^c e^V
{\bar \Phi}^c + {\bar \Phi} e^{-V} \Phi \right)
\right.\nonumber\\&&\left. + \int{d^2\theta} \left( {\Phi}^c
(\partial_5 -{1\over {\sqrt 2}} \Sigma) \Phi + {\rm H. C.}
\right)\right]~.~\, \label{VD-Lagrangian}
\end{eqnarray}

We introduce the following orbifold projections
  \beqa
Z_5:x_5&\ra&-x_5~,~~~~T_5:x_5\ra x_5+2\pi R_5~,
  \eeqa
and use them to impose the following boundary conditions on vector and
hypermultiplets in terms of the fundamental representation
  \beqa
    V(-x_5)&=&Z_5 V(x_5) Z_5~,~~~
    \Sigma_5(-x_5)=-Z_5 \Sigma_5(x_5) Z_5~,\\
    \Phi(-x_5)&=&\eta_\Phi Z_5 \Phi(x_5) ~,~~~\Phi^c(-x_5)=-\eta_\Phi Z_5 \Phi(x_5)
    ~,
  \eeqa
and
  \beqa
V(x_5+2\pi R_5)&=&T_5 V(x_5) T_5~,~~ \Sigma_5(x_5+2\pi R_5)=T_5
\Sigma_5(x_5) T_5~,\\
\Phi(x_5+2\pi R_5)&=&\zeta_{\Phi} T_5 \Phi(x_5)~,~~\Phi^c(x_5+2\pi
R)=\zeta_{\Phi}T_5 \Phi(x_5) ~, \eeqa with
$\eta_\Phi=\pm1$, and $\zeta_\Phi=\pm1$.
The 5D ${\cal N}=1$ supersymmetry, which corresponds to 4D ${\cal N}=2$ SUSY,
reduces to 4D ${\cal N}=1$ supersymmetry after the $Z_5$ projection.

   It is well known that we can have different gauge symmetries at the
two fixed points by assigning different boundary conditions. We can
rewrite $(Z_5,T_5)$ in terms of $(Z_5,Z_6)$ by introducing the
transformation \beqa Z_6=T_5Z_5~, \eeqa which gives \beqa Z_6:y+\pi R\ra
-y+\pi R~. \eeqa Then the massless zero modes can preserve different
gauge symmetries which are obtained by assigning proper $(Z_5,Z_6)$ boundary
conditions to the two fixed points.

In our setup, as a warm up, we consider a $SU(7)$ gauge symmetry in the 5D bulk of
${\cal M}_4\tm S^1/Z^2$.
This implies the following different symmetry breaking possibilities.
\begin{itemize}

\item Case I:
\beqa
Z_5 = I_{4,-3}~, ~~~ Z_6 = I_{1,-6}~,
 \eeqa
where $I_{a,-b}$ denotes the diagonal matrix with the first $a$ entries $1$ and the last $b$ entries $-1$.
These boundary conditions break the $SU(7)$ gauge symmetry down to
$SU(4)_W\tm SU(3)_C\tm U(1)_{B-L}$ at the fixed point $y=0$, to
$SU(6)\tm U(1)\tm U(1)_{X}$ at the fixed point $y=\pi R_5$, and
preserve $SU(3)_C\tm SU(3)_L\tm U(1) \tm U(1)_X$ in the low energy
4D theory.

\item Case II:
\beqa
Z_5 = I_{4,-3}~, ~~~ Z_6 = I_{2,-5}~,
 \eeqa
which break the gauge symmetry $SU(7)$ to
$SU(4)_w\tm SU(3)_C\tm U(1)_{B-L}$ at the fixed point $y=0$, to
$SU(5)\tm SU(2)\tm U(1)_{X}$ at the fixed point $y=\pi R_5$, and
preserve $SU(3)_C\tm SU(2)_L\tm SU(2)_R\tm U(1)_{B-L}\tm U(1)_X$ in
the low energy 4D theory.

\item Case III:
\beqa
Z_5 =  I_{4,-3}~, ~~~ Z_6 = I_{3,-4}~,
 \eeqa
which break $SU(7)$ to
$SU(4)_w\tm SU(3)_C\tm U(1)_{B-L}$ at the fixed point $y=0$, to
$SU(3)\tm SU(4)_c\tm U(1)_{X}$ at the fixed point $y=\pi R_5$, and
preserve $SU(3)_C\tm SU(3)_L \tm U(1) \tm U(1)_X$ in the low energy
4D theory.

\item Case IV:
\beqa
Z_5 = I_{1,-6}~, ~~~ Z_6 = I_{2,-5}~,
\eeqa
which break $SU(7)$ to
$SU(6)\tm U(1) $ at the fixed point $y=0$, to $SU(5)\tm SU(2)\tm
U(1)_{X}$ at the fixed point $y=\pi R_5$ which preserves $SU(5) \tm
U(1) \tm U(1)_X$ in the low energy 4D theory.

\item Case V:
\beqa
Z_5 = I_{1,-6}~, ~~~ Z_6 = I_{4,-3}~,
\eeqa
which break $SU(7)$ to
$SU(6)\tm U(1) $ at the fixed point $y=0$, to $SU(3)_C\tm
SU(4)\tm U(1)_{X}$ at the fixed point $y=\pi R_5$, and preserve
$SU(3)_C \tm SU(3)_L \tm U(1)$ in the low energy 4D
theory.

\end{itemize}
We will not discuss these various symmetry breaking chains in detail,
we simply note that several interesting low energy theories can be
embedded into a 5D $SU(7)$ gauge theory. To construct a realistic theory,
we must also introduce the proper matter content. The simplest
possibility to introduce matter in this scenario is to localize it at
the fixed point branes and fitting it into multiplets of the corresponding
gauge symmetry preserved in the given brane.
Bulk fermions which are $SU(7)$ invariant are possible in case
$Z_5$ or $Z_6$ is trivial. For most general boundary
conditions, bulk fermions do not always lead to realistic
low energy matter content. In our case, the motivation for the $SU(7)$
gauge symmetry is the unification of $SU(3)_C\tm SU(4)_w\tm
U(1)_{B-L}$. Thus, we will discuss in detail this symmetry breaking
chain and the matter content of this scenario.

The compactification of gauge symmetry in flat and warped extra
dimensions share many common features.  Consequently, we will concentrate
on the orbifold breaking of $SU(7)$ in a warped extra dimension and
discuss its AdS/CFT interpretation. The flat extra dimension
results can be obtained by taking the AdS curvature radius to infinity.

\section{SUSY $SU(7)$ Unification in Warped Extra Dimension}
\label{sec-2}
We consider the AdS$_5$ space warped on $S^1/Z_2$ with $SU(7)$
bulk gauge symmetry. The AdS metric can be written as
 \beqa
\label{RS} ds^2=e^{-2\sigma}\eta_{\mu\nu}dx^\mu dx^\nu+dy^2~,
\eeqa
where $\sigma=k|y|$, $1/k$ is the AdS curvature radius, and $y$ is the
coordinate in the extra dimension with the range $0\leq y\leq\pi
R$. Here we assume that the warp factor $e^{-k\pi R}$ scales ultra-violet
(UV) masses to TeV.

As noted in~\cite{susyads0}, in AdS space the different fields
within the same supersymmetric multiplets acquire different masses.
The action for bulk vector multiplets $(V_M,\la^i,\Sigma)$ and
hypermultiplets $(H^i,\Psi)$ can be written as~\cite{susy-ads,
susy1-pomarol-tony,susy2-pomarol-tony} \beqa
 S_5=-\frac{1}{2}
\int d^4x\int dy\sqrt{-g}&&\Bigg[\frac{1}{2g^2_5}F^2_{MN}+
\left(\partial_M\Sigma\right)^2+i\bar{\lambda^i}\gamma^MD_M\lambda^i
+m^2_\Sigma\Sigma^2~,\nn\\+&&
im_\lambda\bar{\lambda^i}(\sigma_3)^{ij}\lambda^j+\left|\partial_M
H^i \right|^2+ i\bar{\Psi}\gamma^MD_M\Psi
     +m^2_{H^i}|H^i|^2~,\nn\\+&& im_\Psi\bar{\Psi}\Psi \Bigg]\, , \eeqa
with supersymmetry preserving mass terms for vector multiplets \beqa
 m^2_\Sigma&=&-4k^2+2\sigma^{\prime\prime}\, , \\
       m_\lambda&=&\frac{1}{2}\sigma^\prime\, ,
\eeqa and for hypermultiplets \beqa
  m^2_{H^{1,2}}&=&(c^2\pm c-\frac{15}{4})k^2
       +\left(\frac{3}{2}\mp c\right)\sigma^{\prime\prime}\, , \\
       m_\Psi&=&c\sigma^\prime\, .
\eeqa We introduce the generic notation
\beqa
m_\phi^2&=& ak^2+b\sigma^{\prime\pr}~, \\
m_\psi&=&c\sigma^\pr~,
\eeqa
for the AdS mass terms of bosons ($\phi$) and fermions ($\psi$)
with
\beqa
\sigma^\pr&=&\f{d\sigma}{dy}=k\epsilon(y)~,\\
\sigma^{\pr\pr}&=&2k\[\delta(y)-\delta(y-\pi R)\]~,
\eeqa
where the step function is defined as $\epsilon(y)= +1~(-1)$ for positive (negative) $y$.
With this notation, we can
parametrize the bulk mass terms for vector multiplets as
\beqa
a=-4~, ~~~ b=2~, ~~~ c=\f{1}{2}~,
\eeqa
and for hypermultiplets as
\beqa
a=c^2\pm c-\f{15}{4}~, ~~~ b=\f{3}{2}\mp c~.
\eeqa
 The parameter $c$ controls the zero mode wave
function profiles~\cite{susy1-pomarol-tony,tony-les}.  When $c>1/2$,
the massless modes will be localized towards the $y=0$ (UV) brane.
The larger the value of $c$ the stronger is the localization.
On the other hand, when $c<1/2$, the zero modes will be localized
towards the $y=\pi R$ (IR) boundary. Kaluza-Klein (KK) modes
localized near the IR brane, according to the AdS/CFT dictionary,
correspond dominantly to CFT bound states.

  The $SU(7)$ gauge symmetry can be broken into $SU(3)_C\tm SU(4)_W\tm U(1)_{B-L}$
by the Higgs mechanism or by boundary conditions.
The spontaneous breaking of $SU(7)$ will lead to the
doublet-triplet (D-T) splitting problem. Thus, it is advantageous to
consider the breaking of the gauge symmetry via boundary conditions
which elegantly eliminates the D-T splitting problem.
We chose the following boundary conditions in terms of $Z_5$ and
$Z_6$ parity
 \beqa
Z_5&=&(+1~,+1~,+1~,-1~,-1~,-1~,-1~)~,\\
Z_6&=&(+1~,+1~,+1~,+1~,+1~,+1~,+1~)~, \eeqa
which break $SU(7)$
to $SU(3)_C\tm SU(4)_W \tm U(1)_{B-L}$ at the fixed point $y=0$.
  The parity assignments of $SU(7)$ vector supermultiplets
in terms of $(Z_5,Z_6)$ are
\beqa V^g({\bf 48}) &=&V^{++}_{({\bf
8,1})_0}\oplus V^{++}_{({\bf 1,15})_0} \oplus V^{++}_{({\bf
1,1})_0}\oplus V^{-+}_{({\bf 3,\bar{4}})_{7/3}}\oplus
V^{-+}_{({\bf \bar{3},4})_{-7/3}}~,  \nn\\
\Sigma^g({\bf 48})&=&\Sigma^{--}_{({\bf 8,1})_0}\oplus
\Sigma^{--}_{({\bf 1,15})_0} \oplus \Sigma^{--}_{({\bf
1,1})_0}\oplus \Sigma^{+-}_{({\bf 3,\bar{4}})_{7/3}}\oplus
\Sigma^{+-}_{({\bf \bar{3},4})_{-7/3}}~, \eeqa
where the lower indices show the $SU(3)_C\tm SU(4)_W \tm U(1)_{B-L}$
quantum numbers.
After KK decomposition, only the ${\cal N}=1$ SUSY
$SU(3)_C\tm SU(4)_W \tm U(1)_{B-L}$ components
of the vector multiplet $V^g$ have zero modes. Kaluza-Klein
modes which have warped masses of order $M_{UV} e^{-k\pi R}\sim {\rm TeV}$ are
localized towards the symmetry preserving IR brane, so they are
approximately $SU(7)$ symmetric.

It is also possible to break the gauge symmetry on the $y=\pi R$
brane (by interchanging $Z_5$ and $Z_6$) which will have a different
4D CFT dual description in contrast to the previous case. If
$SU(7)$ breaks to $SU(3)\tm SU(4)_W\tm U(1)_{B-L}$ on the IR
brane, the dual description is a technicolor-like theory in which
$SU(7)$ is broken to $SU(3)\tm SU(4)_W\tm U(1)_{B-L}$ by strong dynamics at
the TeV scale. In case the gauge symmetry is broken on the
UV brane, the dual descriptions is a theory with $SU(7)$ global
symmetry and a weakly interacting $SU(3)\tm SU(4)_W\tm
U(1)_{B-L}$ gauge group at the UV scale. The IR brane with a
spontaneously broken conformal symmetry respects the $SU(7)$ gauge
group. In this scenario, the $SU(7)$ gauge symmetry is composite(emergent) which
is similar to the rishon model~\cite{seiberg-rishon}. In order to
reproduce the correct Weinberg angle $\sin^2\theta_W$, it is in general not
advantageous to break the GUT symmetry on the IR brane because, from
the AdS/CFT correspondence, the running of the gauge couplings is $SU(7)$
invariant at the TeV scale.

 It is possible to strictly localize matter on the UV brane
that preserves the $SU(3)\tm SU(4)_W\tm U(1)_{B-L}$ gauge symmetry.
However, in this case we will not get a prediction of the weak mixing
angle because we lack the absolute normalization factor of
the hypercharge of the SM particles.\footnote{The
relative normalization within the matter sector is determined by
anomaly cancellation requirements.} Thus it is preferable to place
matter in $SU(7)$ multiplets into the 5D bulk so it can be
approximately localized towards the UV brane by
introducing bulk mass terms. In this case, the $U(1)_{B-L}$ charges
of matter are quantized according to $SU(7)$ multiplet assignments
and we could understand the observed electric charge
quantization of the Universe.

 We arrange quark supermultiplets into ${\bf 28,\overline{28}}$ symmetric $SU(7)$ representations and lepton multiplets into
${\bf 7,\bar{7}}$ representations
\beqa QX({\bf 28})_a&=&\(\bea{cc}({\bf {6},1})_{8/3}&({\bf
3,4})_{1/3}\\({\bf
\bf{3},\bf{4}})_{1/3}&({\bf 1,10})_{-2}\eea\)~,\\
\overline{QX}({\bf \overline{28}})_a&=& \(\bea{cc}({\bf
\bar{6},1})_{-8/3}&({\bf \bar{3},\bar{4}})_{-1/3}\\({\bf
\bar{3},\bar{4}})_{-1/3}&({\bf 1,\overline{10}})_{2}\eea\)~,\\
LX({\bf 7})_a&=&{\bf (3,1)_{{4}/{3}}\oplus(1,4)_{-1}}~,\\
\overline{LX}({\bf \bar{7}})_a&=&{\bf (\bar{3},1)_{-4/3}\oplus
(1,\bar{4})_1}~, \eeqa with the subscript $a$ being the family
index.

To obtain zero modes for chiral quark and lepton multiplets, we
assign the following $(Z_5,Z_6)$ parities to them
\beqa
QX({\bf 28})_a&=&{\bf ({6},1)_{8/3}^{-,+}\oplus (1,10)_{-2}^{-,+}\oplus (3,4)_{1/3}^{+,+}}~,\\
\overline{QX}({\bf \overline{28}})_a&=&{\bf (\bar{6},1)_{-8/3}^{-,+}\oplus (1,\overline{10})_{2}^{-,+}\oplus (\bar{3},\bar{4})_{-1/3}^{+,+}}~,\\
LX({\bf 7})_a&=&{\bf (3,1)^{-,+}_{{4}/{3}}\oplus(1,4)^{+,+}_{-1}}~,\\
\overline{LX}({\bf \bar{7}})_a&=&{\bf (\bar{3},1)^{-,+}_{-4/3}\oplus
(1,\bar{4})^{+,+}_1}~. \eeqa
Parity assignments for the conjugate fields $\Phi^c$ are
opposite to those for $\Phi$.

Because of the unification of $SU(3)_C\tm SU(4)_W\tm U(1)_{B-L}$ into $SU(7)$,
we can determine the normalization of the $U(1)_{B-L}$ charge based on the
the matter sector charge assignments in the fundamental representation of $SU(7)$
  \beqa
Q_{B-L}=(~4/3,~4/3,~4/3,-1,-1,-1,-1)~.
  \eeqa
Then from the relation of the gauge couplings
 \beqa g_{B-L}\f{Q_{B-L}}{2}=g_7
T^{B-L}~, \eeqa
we obtain
\beqa g_{B-L}=\sqrt{\f{3}{14}}g_7~. \eeqa
Here we normalize the $SU(7)$ generator as
\beqa
Tr(T^aT^b)=\f{1}{2}\delta^{ab}~.
\eeqa
Thus the tree-level weak mixing angle can be predicted to be
\beqa
\sin^2\theta_W\equiv\f{g_Y^2}{g_Y^2+g_{L}^2}=\f{3}{20}=0.15~.
\eeqa

In previous SUSY $SU(7)$ unification scenario with quark contents fitting in ${\bf 28},{\bf \overline{28}}$ dimensional representations,
there is no ordinary proton decay problem related to heavy gauge boson exchanges (D-type operators) and dimension-five operators which can be seen from
the charge assignments of the $SU(7)$ matter multiplets. Contributions from dimension
four operators of the form
$\la_{ijk}(QX)_i(\overline{LX})_j(\overline{LX})_k+
\tl{\la}_{ijk}(\overline{QX})_i({LX})_j({LX})_k$
can be forbidden by R-parity. However, it is also possible to fit the quark sectors in ${\bf 21},{\bf \overline{21}}$ dimensional representations of SU(7) instead of ${\bf 28},{\bf \overline{28}}$ dimensional representations.
Then dangerous IR-brane localized dimension five F-type operators
of the form \beqas
{\cal L}=\int d^2\theta\f{1}{M}
  \!\!\! &[& \!\! \la_{1ijkl}(QX)_i(QX)_j(QX)_k(LX)_l+ \la_{2ijkl}(\overline{QX})_i(\overline{QX})_j
                 (\overline{QX})_k(\overline{LX})_l],
 \eeqas
can be introduced. Such dimension five operators can arise from a diagram involving the coupling of
matter to the ${\bf 35},\overline{\bf 35}$ Higgs multiplet and the insertion of
$\mu$-term like mass terms for such Higgs fields. If matter is localized towards the IR brane, then the
suppression scale $M$ is of order TeV and this results in rapid proton
decay. Since the profile of zero modes for bulk
matter with $c\gtrsim1/2$ is
\beqa \phi_+^{(0)}\sim e^{-(c-\f{1}{2})ky}~, \eeqa
we could assign bulk mass terms to matter with $c\gtrsim1/2$
to suppress the decay rates.
Then for an IR brane localized dimension five operator,
we require
\beqa \f{1}{(\rm{TeV})}e^{-\sum\limits_{i}(c_i-\f{1}{2})k\pi
R} \lesssim\f{10^{-8}}{M_{Pl}}\approx\f{e^{-3k\pi R/2}}{(\rm{TeV})}~,
\label{Proton-Bound}
\eeqa
which
satisfies proton decay bounds~\cite{Harnik:2004yp}.
However such requirements will lead to difficulty in giving natural Yukawa couplings. So we consider only the case with quark sector fitting in ${\bf 28}$ and ${\bf \overline{28}}$ dimensional representations.

There are several ways to introduce Yukawa couplings.
Orbifold GUTs are well known to solve the
D-T splitting problem by assigning appropriate
boundary conditions to bulk Higgs fields. Thus, it is also possible
to introduce bulk Higgs fields in our scenario. Since
 \beqa {\bf 28}\otimes\overline{\bf 28}&=&{\bf
1}\oplus{\bf 48}\oplus{\bf
735}~,\\
{\bf 7}\otimes\overline{\bf 7}&=&{\bf 1}\oplus{\bf 48}~,\\
\overline{\bf 7}\otimes\overline{\bf 7}&=&\overline{\bf
21}\oplus\overline{\bf 28}~,
 \eeqa
we can introduce bulk Higgses $\Sigma,\tl{\Sigma}$ in the $SU(7)$ adjoint
representation ${\bf 48}$, $\Delta_1,\Delta_2$ in $SU(7)$ symmetric
representations ${\bf 28},\overline{\bf 28}$, and an $SU(7)$ singlet
Higgs $S$ to construct $SU(7)$ gauge invariant Yukawa couplings. We
impose the following boundary conditions on the bulk Higgs fields
\beqa \Sigma,\tl{\Sigma}({\bf 48}) &=&({\bf 8,1})_0^{-,+}\oplus({\bf
1,15})_0^{+,+} \oplus ({\bf 1,1})_0^{-,+}\oplus ({\bf
3,\bar{4}})_{7/3}^{-,+}\oplus
({\bf \bar{3},4})_{-7/3}^{-,+}~, \\
\Delta_1({\bf 28})&=&{\bf ({6},1)_{8/3}^{-,+}\oplus (1,{10})_{-2}^{+,+}\oplus ({3},{4})_{1/3}^{-,+}}~,\\
\Delta_2(\overline{\bf 28})&=&{\bf (\overline{6},1)_{-8/3}^{-,+}\oplus (1,\overline{10})_{2}^{+,+}\oplus (\bar{3},\bar{4})_{-1/3}^{-,+}}~,\\
S({\bf 1})&=&({\bf 1,1})_0^{+,+}~. \eeqa In the orbifold
projection above, we choose the most general boundary
conditions~\cite{csaki1,csaki2,csaki3} to eliminate unwanted zero
modes. The results can be obtained from naive orbifolding by
introducing the relevant heavy brane mass terms (on the UV brane) to
change the Neumann boundary conditions to Dirichlet ones. Then
the surviving zero modes give the Higgs content required in a
4D SUSY $SU(3)_C\tm SU(4)_W\tm U(1)_{B-L}$ theory\footnote{We can also
eliminate the bulk singlet Higgs field $S$ and choose the boundary
conditions so that the $SU(3)_C\tm SU(4)_W\tm U(1)_{B-L}$ singlet
comes from (projections of) bulk Higgs hypermultiplets $\Sigma({\bf
48})$. An additional $\Sigma_2({\bf 1,15})_0$ from $\tl{\Sigma}$ is
required to break $SU(4)_W\tm U(1)_{B-L}$ to $SU(2)_L\tm
SU(2)_R\tm U(1)_Z\tm U(1)_{B-L}$.}. Bulk Yukawa couplings can be
introduced as
\beqa S_{5D}&=&\int d^4x\int dy \sqrt{-g}
\int{d^2\theta}\sum\limits_{i=1,2,3}\Big{(}\tl{y}_{1ij}^{QX}
(QX)^i\Sigma(\overline{QX})^j+\tl{y}_{1ij}^{LX}(LX)^i\Sigma(\overline{LX})^j\nn\\
&&+\tl{y}_{2ij}^{QX}(QX)^iS(\overline{QX})^j+\tl{y}_{2ij}^{LX}(LX)^iS(\overline{LX})^j+\tl{y}_{3ij}^{LX}(\overline{LX})^i\Delta_1(\overline{LX})^j\nn\\
&&+\tl{y}_{4ij}^{LX}({LX})^i\Delta_2({LX})^j\Big{)}~. \eeqa
Then at low energies, after the heavy KK modes are projected out, the effective
4D Yukawa couplings are
\beqa S_{4D}=\int d^4x
\int{d^2\theta}&&\sum\limits_{i=1,2,3}\(y_{1ij}^QQ_L^i\Sigma_1(Q_L^c)^j+y_{1ij}^LL_L^i\Sigma_1(L_L^c)^j
+y_{2ij}^QQ_L^iS(Q_L^c)^j\.\nn\\&&+\left.y_{2ij}^LL_L^iS(L_L^c)^j+y_{ij}^{N^c}(L_L^c)^i\Delta
(L_L^c)^j+y_{ij}^{N}(L_L)^i\overline\Delta (L_L)^j\)~.
\eeqa
Here we
denote the $SU(3)\tm SU(4)_W\tm U(1)_{B-L}$ multiplet
${\bf(3,4)_{1/3}}$ by $Q_L$, the ${\bf (\bar{3},\bar{4})_{-1/3}}$
by $Q_L^c$, the ${\bf(1,4)_{-1}}$ by $L_L$, the ${\bf (1,\bar{4})_{1}}$
by $L_L^c$, the $({\bf 1,15})_0$ by $\Sigma_1$, the $({\bf 1,10})_2$ by
$\Delta$, the $({\bf 1,\overline{10}})_{-2}$ by $\overline{\Delta}$,
and the ${\bf (1,1)_0}$ by $S$. As indicated in Ref.~\cite{fei1}, such
Yukawa interactions are necessary in a 4D theory to give acceptable
low energy spectra. The SM fermionic masses and mixing hierarchy,
which is related to the coefficients of the 4D
Yukawa couplings, can be understood from the wave
function profile overlaps~\cite{susy1-pomarol-tony,AJ}. The profile of the
bulk Higgs fields can also be determined from their bulk mass terms.
In our scenario, bulk Higgses other than the ${\bf 35}$ multiplet
are not responsible for proton decay and can be localized anywhere
(such as towards the IR
brane to generate enough hierarchy). For simplicity, we can set the
zero modes of bulk Higgs profiles to be flat with the mass terms
$c_\Sigma=c_{\tl{\Sigma}}=c_S=c_{\Delta_i}=c_{H}=1/2$. The low
energy Yukawa coupling coefficients that appeared in previous expressions
are of order \beqa y_{ij}\sim 4\pi
\tl{y}^{QX,LX}_{ij}\sqrt{k}\(\prod_i\sqrt{\f{1-2c_i}{e^{-2(c_i-\f{1}{2})k\pi
R}-1}}\) e^{-(c_{Xi}+c_{\bar{X}i}+c_{H}-\f{3}{2})k\pi R}~, \eeqa
which can generate the required mass hierarchy and CKM mixing by the
Froggatt-Nielson mechanism~\cite{FN}. We can, for example, chose
\beqa
c_1^{QX}&=& c_1^{\overline{QX}}= \f{1}{2}+\f{1}{8},~~~c_1^{LX}=
c_1^{\overline{LX}}=\f{1}{2}+\f{1}{16}~,\\c_2^{QX}&=& c_2^{\overline{QX}}=
\f{1}{2}+\f{1}{16},~~c_2^{LX}= c_2^{\overline{LX}}=
\f{1}{2}+\f{1}{32}~,\\c_3^{QX}&=& c_3^{\overline{QX}}= \f{1}{2},~~~~~~~~~
c_3^{LX}= c_3^{\overline{LX}}=\f{1}{2}~.
\eeqa
Here we use the
fact that $e^{-k\pi R} \simeq ({\rm TeV})/k\simeq {\cal
O}(10^{-16})$ and $\tl{y}^{QX,LX}_{ij}\sqrt{k}\sim {\cal O}(1)$. Besides,
it is obvious from the charge assignment in the matter sector
that there are no unwanted mass relations in our scenario, such as
$m_{\mu}:m_{e}=m_s:m_d$, that appear in an $SU(5)$ GUT.

\section{Gauge Coupling Unification in SUSY SU(7) Unification}
\label{sec-3} The Lagrangian relevant for the low energy gauge
interactions has the following form
\beqa\label{gauge-beta} S=\int d^4x \int\limits_{0}^{\pi R}d y
\sqrt{-g}\[-\f{1}{4g_5^2}F^{aMN}F_{MN}^a-\delta(y)\f{1}{4g_{0}^2}F_{\mu\nu}^aF^{a\mu\nu}-\delta(y-\pi
R)\f{1}{4g_{\pi}^2}F_{\mu\nu}^aF^{a\mu\nu}\]~,\nn \eeqa where $g_5$
is the dimensionful gauge coupling in the 5D bulk, $g_{0}$ and
$g_{\pi}$ are the relevant gauge couplings on the $y=0$ and $y=\pi
R$ brane, respectively. The brane kinetic terms are necessary
counter terms for loop corrections of the gauge field propagator.
In AdS space there are several tree-level mass scales which are related as
\beqa \mu\ll M_{KK}\simeq\f{\pi k}{e^{\pi k R}-1}\ll \f{1}{R} \ll
k\simeq M_*~. \eeqa Thus, the 4D tree-level gauge couplings can be
written as~\cite{nomura2}\beqa \f{1}{g_a^2}(\mu)=\f{\pi
R}{g_5^2}+\f{1}{g_{0}^2}+\f{1}{g_{\pi}^2}+\f{1}{8\pi^2}\tl{\Delta}(\mu,Q)~,
\eeqa where the first three terms contain the tree-level gauge couplings, and
$\tl{\Delta}(\mu,Q)$  represents the one-loop corrections. The explicit
dependence on the subtraction scale cancels that of the running
boundary couplings in such a way that the quantity $g^2_a(\mu)$ is
independent of the renormalization scale. We assume that the bulk
and brane gauge groups become strongly coupled at the 5D
Planck scale $M_{5D}=\Lambda$ with \beqa
\f{1}{g_{0}^2}(\Lambda)\approx\f{1}{g_{\pi}^2}(\Lambda e^{-k \pi
R})\approx\f{1}{16\pi^2}~,~~~~~~\f{\pi
R}{g_5^2}(\Lambda)\approx{\cal O}(1)~. \eeqa The GUT breaking
effects at the fixed points are very small compared to bulk GUT
symmetry preserving effects. Thus, we can split the contributions to the
gauge couplings into symmetry preserving and symmetry breaking pieces
  \beqa \f{1}{g_a^2}(\mu)&=&\f{\pi
R}{g_5^2}(\Lambda)+\f{1}{g_{0}^2}(\Lambda)+\f{1}{g_{\pi}^2}(\Lambda
e^{-k \pi
R})+\f{1}{8\pi^2}\[\tl{\Delta}(\mu,\Lambda)+b_0^a\ln\f{\Lambda}{\mu}+b_{\pi
}^a\ln\f{\Lambda e^{-k \pi R}}{\mu}\]~,\nn\\
&\simeq&({\rm SU(7)~
symmetric})+\f{1}{8\pi^2}\[\tl{\Delta}(\mu,\Lambda)+b_0^a\ln\f{\Lambda}{\mu}+b_{\pi
}^a\ln\f{\Lambda e^{-k \pi R}}{\mu}\]~,\nn\\
&{\equiv}&({\rm SU(7)~
symmetric})+\f{1}{8\pi^2}{\Delta}(\mu,\Lambda)~.
 \eeqa
The general expression for $\Delta(\mu,\Lambda)$ was calculated in~\cite{CKS}.
The contributions from the vector multiplets are \beqa
\left(
\bea{c}\Delta_{U(1)_{B-L}}\\\Delta_{SU(3)_C}\\\Delta_{SU(4)_W}\eea
\right)_V=({\rm SU(7)~
symmetric})+\(\bea{c}0\\-9\\-12\\\eea\)\ln\left(\f{k}{\mu}\right)~,\eeqa with
\beqa
&&T(V_{++})=(~0,~3,~4)~\\
&&T(V_{-+})=(~7,~4,~3)~, \eeqa for $U(1)_{B-L}$, $SU(3)_C$, and $SU(4)_W$,
respectively.
 Here we normalize the $U(1)_{B-L}$
gauge coupling according to $g_{B-L}^2=3g_7^2/14$~.

 We can use the facts $c^{QX}_i,c^{LX}_i\geq1/2$ to simplify
the matter contributions to \beqa
\Delta_M(\mu,k)&=&T(H_{++})\[\ln\left(\f{k}{\mu}\right)-c_H\ln\left(\f{k}{T}\right)\]+c_HT(H_{+-})\ln\left(\f{k}{T}\right)\nn\\
&-&c_HT(H_{-+})\ln\left(\f{k}{T}\right)+T(H_{--})\[\ln\left(\f{k}{\mu}\right)-(1+c_H)\ln\left(\f{k}{T}\right)\]~.
\eeqa
 Thus the
contributions from the bulk matter hypermultiplets are \beqa \left(
\bea{c}\Delta_{U(1)_{B-L}}\\\Delta_{SU(3)_C}\\\Delta_{SU(4)_W}\eea
\right)_H=({\rm SU(7)~
symmetric})+\(\bea{c}\f{12}{7}\\12\\12\\\eea\)\ln\left(\f{k}{\mu}\right) ~,\eeqa
with
 \beqa
&&T(H_{++})|_{H+H^c}^m=\left(~\f{12}{7},~12,~12\right)~,\\
&&T(H_{-+})|_{H+H^c}^m=\left(\f{198}{7},~18,~18\right)~. \eeqa

 The contributions from the bulk Higgs hypermultiplets include two
 ${\bf 48}$ dimensional representations, ${\bf 28}$ and
$\overline{\bf 28}$ dimensional
 representations and possible one singlet.\footnote{ We can also add
Higgs fields in the ${\bf 35}$ and $\overline{\bf 35}$ representations
with flat profiles and impose the following boundary conditions
\beqa {\bf 35} &=&({\bf 1,1})_4^{-,+}\oplus({\bf
1,\bar{4}})_{-3}^{+,+} \oplus ({\bf \bar{3}, {4}})_{5/3}^{+,+}\oplus
({\bf {3},{6}})_{-2/3}^{-,+}~,\\
\overline{\bf 35} &=&({\bf 1,1})_{-4}^{-,+}\oplus({\bf 1,
{4}})_3^{+,+} \oplus ({\bf {3}, \bar{4}})_{-5/3}^{+,+}\oplus ({\bf
\bar{3},\bar{6}})_{2/3}^{-,+}~. \eeqa Then the beta functions receive
additional contributions:
\beqa
&&T(H_{++})|_{H+H^c}^h=\left(~\f{52}{7},~4,~4\right)~,\\
&&T(H_{-+})|_{H+H^c}^h=\left(\f{18}{7},~6,~6\right)~. \eeqa }
The contributions from the bulk Higgs hypermultiplets are
 \beqa
\left(
\bea{c}\Delta_{U(1)_{B-L}}\\\Delta_{SU(3)_C}\\\Delta_{SU(4)_W}\eea
\right)_M=({\rm SU(7)~
symmetric})+\(\bea{c}\f{30}{7}\\~0\\~14\\\eea\)\ln\left(\f{k}{\mu}\right)~,\eeqa
with
 \beqa
&&T(H_{++})|_{H+H^c}^h=\left(~\f{30}{7},~0,~14\right)~,\\
&&T(H_{-+})|_{H+H^c}^h=\left(\f{131}{7},~23,~9\right)~. \eeqa
Thus, the total contribution to the RGE running of the three gauge
couplings are \beqa \f{1}{g_a^2}=({\rm SU(7)~
symmetric})+\f{1}{8\pi^2}\Delta_a~, \eeqa with \beqa \left(
\bea{c}\Delta_{U(1)_{B-L}}\\\Delta_{SU(3)_C}\\\Delta_{SU(4)_W}\eea
\right)=({\rm SU(7)~
symmetric})+\(\bea{c}~6\\~3\\~14\\\eea\)\ln\left(\f{k}{\mu}\right)~.\eeqa

We summarize the supermultiplets in SUSY SU(7) GUT model that contribute to running of the three gauge couplings upon $M_{\tl U}$ as follows:
\begin{itemize}
\item  Gauge:  $V^g(\bf 48),\Sigma^g({\bf 48})$.

\item  Matter: $QX_a({\bf 28}),\overline{QX}_a({\bf \overline{28}}),LX({\bf 7}),\overline{LX}_a({\bf \bar{7}})$~~($a=1,2,3$).

\item  Higgs:  $\Sigma({\bf 48}),\tl{\Sigma}({\bf 48}),\Delta_1({\bf 28}),\Delta_2({\bf 28})$.

\end{itemize}

We can also consider the following symmetry breaking chain
for the partial unification $SU(4)_W\tm U(1)_{B-L}$:
\beqa
SU(4)_W\tm U(1)_{B-L}\ra
SU(2)_L\tm SU(2)_R\tm U(1)_Z\tm U(1)_{B-L}\ra SU(2)_L \tm
U(1)_Y~.\nn \eeqa
Detailed discussions on this symmetry breaking
chain can be found in our previous work~\cite{fei1}.

Assuming that the left-right scale, which is typically the $SU(2)_R$ gauge
boson mass scale $M_R$, is higher than that of the soft SUSY
mass parameters
$M_S$, the RG running of the gauge couplings below the $SU(4)_W\tm
U(1)_{B-L}$ partial unification scale $M_{\tl{U}}$ is calculated as follows.
\begin{itemize}
\item

For $M_Z<E<M_S$, the $U(1)_Y, SU(2)_L$, and $SU(3)_C$ beta-functions are
given by the two Higgs-doublet extension of the SM \beqa
(b_1,b_2,b_3)=\(~7,-3,-7\) \, . \eeqa

\item
For $M_S<E<M_R$, the $U(1)_Y,SU(2)_L$, and $SU(3)_C$ beta-functions are
given by \beqa (b_1,b_2,b_3)=\(12,~2,-3\) \, . \eeqa

\item For $M_R<E<M_{\tl{U}}$, the ${\sqrt{2}}U(1)_Z,\sqrt{\f{14}{3}}U(1)_{B-L},SU(2)_L=SU(2)_R$, and $SU(3)_C$ beta functions
      are given by
\beqa (b_0,b_1,b_2,b_3)=\(22,~6,~16,~3\)\, . \eeqa

\end{itemize}
In our calculation the mirror fermions are fitted into $SU(4)_W$
multiplets and acquire masses of order $M_R$.
The $\tl{\Sigma}({\bf 15})$ Higgs
fields decouple at scales below $M_{\tl{U}}$~\cite{fei1}.
We can calculate the $SU(7)$ unification scale when we know the
$SU(4)_W\tm U(1)_{B-L}$ partial unification scale $M_{\tl{U}}$, which can
be determined from the coupling of $U(1)_Z$ at $M_R$. Here we simply
set $M_{\tl{U}}$ as a free parameter. At the weak scale our inputs are~\cite{PDG}
 \beqa
  M_Z&=&91.1876\pm0.0021 ~,~\,\\
  \sin^2\theta_W(M_Z)&=&0.2312\pm 0.0002 ~,~\,\\
  \alpha^{-1}_{em}(M_Z)&=&127.906\pm 0.019 ~,~\,\\
  \alpha_3(M_z)&=&0.1187\pm 0.0020 \, ,
 \eeqa
which fix the numerical values of the standard $U(1)_Y$ and
$SU(2)_L$ couplings at the weak scale \beqa
\alpha_1(M_Z)&=&\f{\alpha_{em}(M_Z)}{\cos^2\theta_W}=(98.3341)^{-1}~,\\
\alpha_2(M_Z)&=&\f{\alpha_{em}(M_Z)}{\sin^2\theta_W}=(29.5718)^{-1}~.
\eeqa

 The RGE running of the gauge couplings reads
\beqa \f{d~\alpha_i}{d\ln E}=\f{b_i}{2\pi}\alpha_i^2~ \, , \eeqa
where $E$ is the energy scale and $b_i$ are the beta functions.
Our numerical results (See fig.1) show that successful unification of the three
gauge couplings is only possible for small $M_R\lesssim 500$ GeV
and relatively high $M_{\tl{U}}$. For example, if we choose $M_S=200$ GeV,
$M_R=400$ GeV and $M_{\tl U}=2.0\tm 10^6$ GeV, we obtain successful
$SU(7)$ unification at \beqa M_U=9.0\tm 10^6~
{\rm GeV}~,\eeqa and \beqa \f{1}{g_U^2}\simeq 4.65~. \eeqa
 Such low energy $M_R$ may be disfavored by electro-weak\cite{MRbounds1} and flavor precision bounds\cite{MRbounds2}.
In general, with additional matter and Higgs contents (for example, additional bulk ${\bf 7,\overline{7}}$ messenger fields), the low $M_R$ requirement for gauge coupling unifications can be relaxed. Besides, symmetry breaking of $SU(3)\tm SU(4)_W\tm U(1)_{B-L}$ will lead to non-minimal left-right model\cite{fei1}. Thus relatively low $M_R$ can be consistent with flavor precision bounds\cite{MRbounds2}. The choice of $M_{\tl U}=2.0\tm 10^6$ generates a hierarchy between the weak scale and the partial unification scale. Lacking the knowledge of $U(1)_Z$ gauge coupling strength upon SUSY left-right scale, the mild hierarchy between partial unification scale $M_{\tl U}$ and SUSY left-right scale can be the consequences of logarithm running of the various gauge couplings.

   It follows from the AdS/CFT correspondence that the $SU(7)$ unification in RS model are also a successful 4D unification. Our $SU(7)$ model is
vector-like and thus anomaly free. The 5D theory in the bulk is
also anomaly free because the theory on the UV
(which is a $SU(3)\tm SU(4)_W\tm U(1)_{B-L}$ theory~\cite{fei1}) and IR branes
is non-anomalous~\cite{nima3}.

\begin{figure}[htb]
\label{GUT}
\begin{center}
\includegraphics[width=5in]{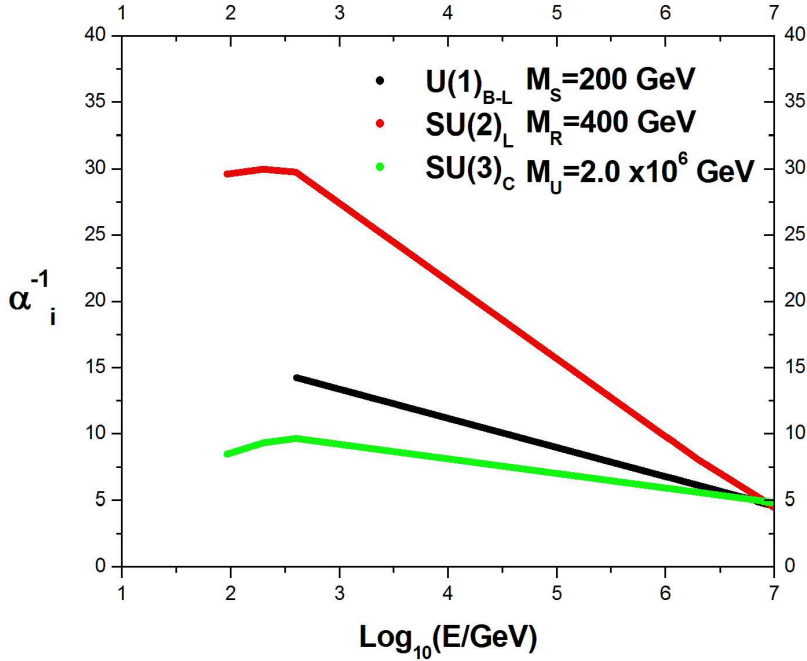}
\end{center}
\caption{\label{spectrum} One loop relative running of the three gauge coupling in SUSY SU(7) GUT model. Here $SU(4)_W$ gauge coupling (upon $M_{\tl U}$) is identified with $SU(2)_L$ gauge coupling. The $U(1)_{B-L}$ gauge coupling strength  at the left-right scale $M_R$ is determined by $U(1)_Y$ and $SU(2)_L$ gauge coupling. Due to the discontinuity between $U(1)_{B-L}$ and $U(1)_Y$ gauge coupling at $M_R$, we do not show the $U(1)_Y$ running below $M_R$ in this figure.}
\end{figure}
\section{Supersymmetry Breaking and Semi-direct Gauge Mediation}
\label{sec-4}
The orbifold projection reduces the 5D ${\cal N}=1$ supersymmetry, which amounts
to 4D ${\cal N}=2$ SUSY, to 4D ${\cal N}=1$ supersymmetry.
We need to break the remaining ${\cal N}=1$
supersymmetry to reproduce the SM matter and gauge content.
One interesting possibility is to use the predictive conformal
supersymmetry breaking proposed for vector-like gauge
theories~\cite{yanagida1,yanagida2}. Conformal supersymmetry breaking
in a vector-like theory can be embedded into a semi-direct gauge
mediation model~\cite{semi-direct} by identifying a subgroup of the
flavor group to be the unifying group of the SM.

\subsection{Supersymmetry Breaking in the Conformal Window}
   The setup of conformal supersymmetry breaking in a vector-like theory
involves an ${\cal N}=1$ $SU(N_c)$ gauge theory with $N_Q<N_c$ quarks
$Q_i,\tl{Q}_i~(i=1,\cdots,N_Q)$ in fundamental and anti-fundamental
representations, and $N_Q\tm N_Q$ gauge singlets $S_i^j$.
Messenger fields $P_a,\tl{P}_a~(a=1,\cdots,N_P)$ with mass
$m$ are also introduced to promote the model to a superconformal
theory. The total number of flavors satisfies $3N_C/2< N_Q+N_P <3N_c$. The
superpotential reads \beqa W=\la Tr(S Q \tl{Q})+mP\tl{P}~, \eeqa
with $Tr(S Q \tl{Q})= S_i^j Q^i \tl{Q}_j$.

When the mass parameter $m$ can be neglected, the theory has a
infrared fixed point. When $S_i^j$ develop vacuum expectation values (VEVs),
$Q_i$ and $\tl{Q}_i$ can be integrated out.
Because $N_Q<N_C$, the theory has a runaway vacuum
when all quark fields are integrated out. Such runaway vacuum can be
stabilized by quantum corrections to the K\"ahler potential and leads to
dynamical supersymmetry breaking. The conformal gauge mediation
model is especially predictive because $m$ is its only free parameter.

\subsection{The AdS/CFT Dual of Seiberg Duality in the Conformal Region and Semi-Direct Gauge Mediation}
 The AdS/CFT correspondence~\cite{maldacena} indicates that the compactification of
Type IIB string theory on $AdS_5\tm S^5$ is dual to ${\cal N}=4$
super Yang-Mills theory.  The duality implies a relation between the AdS radius $R$ and
$g_{YM}^2N=g_sN$: $R^4=4\pi g_s N l_s^4$, in string units $l_s$.
The source of an operator in the CFT sides correspond to the boundary value of a bulk field in gravity side. The generating
function of the conformal theory is identified with the gravitational action in terms of $\phi_0$: \beqa
 \left\langle\exp\left(-\int d^4x \phi_0{\cal O}\right) \right\rangle_{CFT}=\exp(-\Ga[\phi_0])~.
\eeqa
The AdS/CFT correspondence can be extended to tell us that any 5D
gravitational theory on $AdS_5$ is holographically dual to some strongly
coupled, possibly large $N$, 4D CFT~\cite{nima-ads,rattazzi}.
The metric of an $AdS_5$ slice can be written as \beqa
ds^2=\f{L^2}{z^2}\(g_{\mu\nu}dx^\mu dx_\nu +dz^2\)~, \eeqa which is
related to RS metric~\cite{rs} by \beqa
z=Le^{y/L}~,\eeqa with $L=1/k$ being the AdS radius. According to the
AdS/CFT dictionary~\cite{tony-les,perez,pomarol-fermion}, the RG
scale $\mu$ is related to the fifth coordinate by $\mu=1/z$.

We introduce the bulk gauge symmetry $SU(N_F)\tm SU(N_P)\tm
SU(N_Q)\tm U(1)_R$ with $SU(N_P)\tm SU(N_Q)\tm U(1)_R$ being the
global symmetry of the 4D theory. The gauge symmetry $SU(N_F)$ is
broken into $SU(N_C)\tm SU(N_F-N_C)$ at the boundary. The matter
content is $N_F$ chiral multiplets in fundamental
and $N_F$ chiral multiplets in anti-fundamental $SU(N_F)$ representations.
We require the boundary conditions to yield $N_F$ chiral multiplets in both the
fundamental and anti-fundamental representations of $SU(N_C)$ at the UV
brane, and $N_F$ chiral multiplets in both the fundamental and anti-fundamental
representations of $SU(N_F-N_C)$ at the IR brane. These
boundary conditions can be realized by choosing the projection modes
at $y=0$ and $y=\pi R$ as \beqa ~P_1(y=0)=P_2(y=\pi
R)=(\underbrace{~1,\cdots,~1}_{N_c},\underbrace{-1,\cdots,-1}_{N_F-N_c})~.
\eeqa Thus, in terms of $SU(N_c)\tm SU(N_F-N_C)$ quantum numbers,
the field parities and projections are
\beqa Q({N_F})&=&(N_C,1)_{++}\oplus(1,N_F-N_C)_{--}~,\\
      Q^c(\overline{N_F})&=&(\overline{N_C},1)_{--}\oplus(1,\overline{N_F-N_C})_{++}~,\\
      P(N_F)&=&(N_C,1)_{++}\oplus(1,N_F-N_C)_{--}~,\\
      P^c(\overline{N_F})&=&(\overline{N_C},1)_{--}\oplus(1,\overline{N_F-N_C})_{++}~,\\
      S(1)&=&(1,1)_{++}~,~~S^c(1,1)=(1,1)_{--}~.
\eeqa
For $Q(N_F)$ and $P(N_F)$ with $c \gg 1/2$, the
$(N_C,1)$ multiplets (denoted by $Q,P$) are fully localized to the UV brane
while the $(1,\overline{N_F-N_C})$ multiplets (denoted by ${q},{p}$ corresponding
to $Q^c,P^c$ respectively) are strictly localized to the IR
brane. The bulk zero modes localized towards the UV
brane correspond to elementary fields. So, in the conformal
supersymmetry breaking setting, we have the fundamental fields
$Q_i,\tl{Q}_i,P,\tl{P}$ and we can introduce their
interactions on the UV brane \beqa W|_{UV}=\la
Tr(SQ\tl{Q})+mP\tl{P}~. \eeqa  The presence of the additional gauge
symmetry $SU(N_F)$ is required by anomaly matching of
$SU(N_C)$ and $SU(N_F-N_C)$ in the Seiberg duality. Anomaly
matching in the Seiberg duality is equivalent to anomaly inflow of
the Chern-Simmons terms of the 5D bulk, which gives opposite
contributions on the two boundaries~\cite{AB}.

 According to the setup of the conformal supersymmetry breaking scenario, we require
the theory to enter a superconformal region when we can neglect the
masses of $P$ and $\tl{P}$. To ensure that the theory is superconformal
in a certain energy interval, and to be predictive, we need to
determine the exact gauge beta functions.
In the 5D picture, we can
determine the beta-functions by calculating the variation of the gauge
couplings with respect to the fifth dimensional coordinate.
The gauge couplings are obtained by
calculating the correlation functions of the conserved currents.
Then from the 5D gauge coupling running~\cite{CKS}, we can obtain the
dependence on the fifth dimension by replacing $k\pi R$ with
$\ln(z/L)=- \ln(\mu L) $. This way, we obtain the following leading contributions
\beqa
\f{1}{g_a^2}&=&-\f{\ln (\mu L)}{k g_5^2}+\f{\ln(\mu
L)}{8\pi^2}\[\f{3}{2}T_a(V_{++})+\f{3}{2}T_a(V_{+-})-\f{3}{2}T_a(V_{-+})-\f{3}{2}T_a(V_{--})\]\nn~\\
&&- \f{\ln(\mu
L)}{8\pi^2}\[(1-c_H)T_a(H_{++})+c_HT_a(H_{+-})-c_HT_a(H_{-+})
 \right. \nn  \\   && \left.
+(1+c_H)T_a(H_{--})\]~.
\eeqa
 To determine the bulk couplings, we consider the $SU(N_F)$ gauge symmetry on the
UV brane, IR brane and in the bulk. Then, by matching the beta function in the
dual description
\beqa b=\f{8\pi^2}{kg_5^2}-\f{3}{2}N_F=-3N_F~,
\eeqa we obtain
\beqa
\f{8\pi^2}{kg_5^2}=-\f{3}{2}N_F~.
\eeqa

  In our case with a bulk gauge group $SU(N_F)$ and
a gauge group $SU(N_c)$ on the UV and IR branes,
for the $SU(N_c)$ gauge couplings we have
\beqa T(V_{++})=N_c~,~~
T(V_{--})=N_F-N_c~,~~T_a(H_{++})=\f{1}{2}~. \eeqa
The leading contributions are\footnote{The matter contributions are valid
for $c_{++}>1/2$. For $c_{++}\leq1/2$, $1-c_P$ in
front of $N_P$ is replaced by $c_P$.}
\beqa
b_a&=&\f{8\pi^2}{kg_5^2}-\f{3}{2}N_c+\f{3}{2}(N_F-N_c)+(1-C_P)N_P+(1-C_Q)N_Q\\
&=&-3N_c+(1-C_P)N_P+(1-C_Q)N_Q~.
\eeqa
The sub-leading contributions to the gauge couplings depend on $\ln\ln\mu$ and
correct the beta functions with
\beqa
\delta b_a=-\f{1}{\ln(\mu L)}T_a(V_{++})\approx
\f{g_a^2N_c b_a}{8\pi^2 }~.
\eeqa

This expression is valid at two-loop level which we
can see reproducing the NSVZ formula~\cite{NSVZ}
\beqa
\f{dg^2}{d\ln
\mu}=-\f{g^4}{8\pi^2}\f{3T(Ad)-\sum_jT(r_j)(1-\gamma_j)}{1-\f{T(Ad)g^2}{8\pi^2}}~,
\eeqa
by identifying  \beqa \gamma_P=C_P~,~~~\gamma_Q=C_Q~. \eeqa
Via the AdS/CFT correspondence,
the bulk mass is related to the conformal dimension of the operator ${\cal
O}$ that couples to p-forms~\cite{witten}
\beqa (\Delta+p)(\Delta+p-4)=m^2~. \eeqa
Since the anomalous dimensions
 $\gamma_P$ and $\gamma_Q$ are determined by the superconformal invariance of the
boundary, we can obtain the bulk mass terms for the $P$ and $Q$
hypermultiplets. We can obtain the scaling dimension of the 4D
superconformal theory via the R-symmetry charge assignments
\beqa \Delta=\f{3}{2}R_{sc}~. \eeqa The $U(1)_R$ symmetry
of the superconformal theory on the UV brane is determined by the
$a$-maximization technique~\cite{amax}, with $a$ defined by t$^`$Hooft
anomalies of the superconformal R-charge \beqa a=\f{3}{32}(3Tr
R^3-TrR)~, \eeqa and the R-charge being the combination of an arbitrarily
chosen R-charge $R_0$ and other U(1) charges \beqa
R=R_0+\sum\limits_i c_iQ_i~. \eeqa
 This value is the same as the one obtained in~\cite{CGM-09051764}. For
 example with $N_c=4,N_Q=3,N_P=5$, it is~\cite{CGM-09051764}
 \beqa
\Delta_s=1.48~, ~~~\Delta_Q=0.765~.
 \eeqa
In the 4D picture the RG fixed points require
$\gamma_S^*+2\gamma_Q^*=0$ because of the superconformal nature of the theory.

From the AdS/CFT point of view the spontaneous breaking of the CFT originates
from the IR brane. In the limit
$c_Q \gg 1/2$ ($c_{\tl{Q}} \ll -1/2$) the
$q$ and $\tl{q}$ fields are localized to the IR brane, which
means that they are composites in the strongly interacting CFT.
The UV brane interaction can be promoted to a bulk Yukawa coupling
between bulk hypermultiplets $S$ and $Q,\tl{Q}$
 \beqa
S=\int d^4x dy\sqrt{-g}\int d^2\theta \la_b S \tl{Q}Q~,
 \eeqa
which, after projection, will give the IR brane coupling
 \beqa
S=\int d^4x dy\sqrt{-g}\int d^2\theta\delta(y-\pi R)\tl{\la} S
\tl{q}q~.
 \eeqa
Thus, we can anticipate interactions of the form $\tl{\la} S
\tl{q}q$ in the IR brane. If $q$ and $\tl{q}$ are not strongly localized,
they are mixtures of composite and elementary particles. The coupling
of $S$ to $q,\tl{q}$ will also lead to a coupling between $S$
and CFT operators ${\cal O}$ at the boundary.
This can also be seen if we completely
localize $q$ and $\tl{q}$. The hypermultiplet $S$ at the UV boundary is a
source of conformal operators. With $c=1/2$ for $S$
the mixing of CFT states ($SU(N_F-N_c)$ singlets) and $S$ is
marginal.\footnote{The mixing is
important for $|c|\leq1/2$ but marginal for $c=1/2$.}
According to the AdS/CFT interpretation\footnote{From the AdS/CFT
dictionary~\cite{pomarol-fermion} we can see that the operator
${\cal O}$ is dynamical appearing in the low energy
superpotential.}, they correspond to the Seiberg dual
superpotential with the coupling of the form \beqa W=
\tl{\la}S\tl{q}q+\omega S{\cal O}~. \eeqa
 The coefficients $\tl{\la}$ and $\omega$ can be determined
by the AdS/CFT correspondence via
two-point correlation functions. We simply match to the
standard Seiberg dual result giving \beqa
\tl{\la}=\f{1}{\mu}~,~~~~~~~\omega=\la~. \eeqa
Here $\mu$ can be defined in the context of SQCD,
where the beta function coefficients for the magnetic ($\tl{b}$)
and electric ($b$) theories and their respective dynamical transmutation
scales $\tl{\Lambda}$ and ${\Lambda}$ are related as
\beqa
\Lambda^b\bar{\Lambda}^{\tl{b}}=(-1)^{F-N}\mu^{b+\tl{b}}~.
\eeqa

 In the dual description the fields related to $P$ and $\tl{P}$
are integrated out after the RGE running from energies $z_{UV}^{-1}$ to
$z_{IR}^{-1}$ if the
mass parameter satisfies $z_{UV}^{-1}>m>z_{IR}^{-1}$. Thus we
anticipate that $\tl{p}$ and $p$ does not appear as massless fields
on the IR brane.
This can also be understood by observing that adding only the UV
mass terms spoils the zero mode solutions. So the original zero
modes $\tl{p}$ and $p$, which are localized towards the IR brane, are no longer
massless and will not appear in the dual superpotential. This
AdS/CFT interpretation of the Seiberg duality is valid in the IR region
for $3/2N_C<N_F<3N_C$ which are strongly coupled. If the mass
parameter is small, $m<z_{IR}^{-1}$, then it appears as a small perturbation
on the UV brane. We then can promote the mass
parameter $m$ to a bulk field $L$, with $L(z_0)=m$, and introduce
bulk Yukawa couplings between $L$ and the $\tl{P},P$ hypermultiplets.
Similarly to the case of $\tl{Q}$ and $Q$,
the dual description on the IR brane has the
form\footnote{In the presence of $P$ and $\tl{P}$ there are also terms of the
form $(K\tl{q}p+M\tl{p}q)/\mu$, which is similar to the case of $S$
with $\la=0$.}
\beqa W&\sim&\f{1}{\mu}\[S\tl{q}q+ L\tl{p}p\]+
\tl{\omega} \langle {\cal O}_1 \rangle L+\la S{\cal
O}~,\\&\sim&\f{1}{\mu}\[S\tl{q}q+ L\tl{p}p\]+m L+\la S {\cal O}~,
 \eeqa with the coefficients, again, determined by
matching to the Seiberg duality. Here we require the conformal symmetry is spontaneously broken by
$\langle {\cal O}_1 \rangle \neq0$.
After integrating out the fields
$S$ and ${\cal O}$ such that \beqa  S=0~,~~~{\cal O}=-\f{1}{\mu}\tl{q}q~, \eeqa
we can see that the F-term of $L$ \beqa
-F_{L}^\da=m+\f{1}{\mu}\tl{p}p~, \eeqa is non-vanishing (by rank
conditions~\cite{semi-direct}) which indicates that SUSY is broken. It was
pointed out in~\cite{semi-direct} that SUSY breaking by F-term VEVs of
$L$ can cause some problems, such as a low energy Landau pole
and vanishing gaugino masses if we identify the flavor symmetry with the
SM gauge group. Thus it is preferable to study the
case with $z_{UV}^{-1}>m>z_{IR}^{-1}$ where we can integrate out
the fields related to $P$ and $\tl{P}$. Neglecting the additional contributions
from $\tl{P}$ and $P$, the 5D action is~\cite{pomarol}
 \beqa {\cal L}&=&\int d^4\theta
\f{1}{2}(T+T^\da) e^{-(T+T^\da)\sigma}\(S^\da e^{-V}S+S^c e^V
S^{c\da}+(S\leftrightarrow Q,\tl{Q})\)\nn~\\&+&\int d^2\theta e^{-3T
\sigma}S^c\[\pa_5-\f{1}{\sqrt{2}}\chi-(\f{3}{2}-c)T\sigma^\pr\]S+h.c.+(S\leftrightarrow
Q,\tl{Q})~\nn\\&+& W_0\delta(y)+e^{-3T \sigma}W_{\pi R}\delta(y-\pi
R)~,\eeqa
where $T$ is the radion supermultiplet \beqa T=R+iB_5+\theta
\Psi_R^5+\theta^2 F_{\tl{S}}~, \eeqa
$B_5$ is the fifth component of the graviphoton,
$\Psi_R^5$ is the fifth component of the right-handed gravitino,
and $F_{\tl{S}}$ is a complex auxiliary field. After the
lowest component of the radion acquires a VEV, we can re-scale the
fields \beqa (S~,~S^c)\ra \f{e^{k|y|}}{\sqrt{R}} (S~,~S^c)~. \eeqa
Neglecting the gauge sector, for the F-terms of $S$ and $S^c$ we have
 \beqa
-F_{S}^\da&=&\f{e^{-k|y|}}{R}\[-\pa_5+(\f{1}{2}+c_S)k\epsilon(y)\]S^c+\f{e^{-k|y|}}{R}\la_b\tl{Q}Q\\
&+&\delta(y)\la\tl{Q}Q+\delta(y-\pi R)e^{-2k\pi R}\(\f{1}{\mu}\tl{q}q+\la {\cal O}\)~,\nn~\\
-F_{S^c}^\da&=&\f{e^{-k|y|}}{R}\[\pa_5-(\f{1}{2}-c_S)k\epsilon(y)\]S~,
\eeqa while for the $Q$ and $\tl{Q}$ fields \beqa
-F_{Q}^\da&=&\f{e^{-k|y|}}{R}\[-\pa_5+(\f{1}{2}+c_Q)k\epsilon(y)\]Q^c+\f{e^{-k|y|}}{R}\la_bS \tl{Q}\\
&+&\delta(y)\la S\tl{Q}+\delta(y-\pi R)e^{-2k\pi R}\f{1}{\mu} S \tl{q} ~,\nn~\\
-F_{Q^c}^\da&=&\f{e^{-k|y|}}{R}\[\pa_5-(\f{1}{2}-c_Q)k\epsilon(y)\]Q~.
\eeqa

The solutions for $S$, $Q$, and $\tl{Q}$ are
 \beqa S(y)&=&C_Se^{(\f{1}{2}-c_S)k|y|}~,\\
Q(y)&=&C_{Q}e^{(\f{1}{2}-c_Q)k|y|}~,\\
\tl{Q}(y)&=&C_{\tl Q}e^{(\f{1}{2}-c_{\tl Q})k|y|}~,
 \eeqa
with the boundary conditions
\beqa C_S=S~, ~~~ C_Q=Q~, ~~~ Qe^{(\f{1}{2}-c_{Q})k\pi R}=q~,
\eeqa
and $c_S=1/2$ for $S$.
Substituting the previous expressions into the flatness conditions
we can see that, except for the boundary terms, the solutions for $S^c$ and
$Q^c$ are
 \beqa S^c(y)&=&\f{\la_b C_Q C_{\tl Q}}{k(c_Q+c_{\tl Q})}\epsilon(y)e^{(\f{1}{2}-c_S-c_Q-c_{\tl Q})k|y|}~,\\
Q^c(y)&=&\f{\la_b C_S C_{\tl Q}}{k(c_S+c_{\tl
Q})}\epsilon(y)e^{(\f{1}{2}-c_S-c_Q-c_{\tl Q})k|y|}~~.
 \eeqa

The boundary conditions determine the SUSY relations
\beqa
S^c(y=0)&=&\la \tl{Q}Q~, ~~~ S^c(y=\pi R)=\f{1}{\mu}{\tl q}q+\la {\cal
O}\\Q^c(y=0)&=&\la S \tl{Q}~, ~~~ Q^c(y=\pi R)=\f{1}{\mu}S {\tl q}~.
 \eeqa
Substituting back into the previous solutions, we find that the
$F_{S}^{\da}$ and $F_Q^\da$ flatness conditions cannot be satisfied at
the same time. So supersymmetry is broken in this scenario. This
conclusion agrees with the conjecture of~\cite{yanagida2} for a
vanishing $S$ VEV. The non-vanishing
F-term VEV of $S$, which has an R-charge $2N_c/N_Q-2\neq 0$, breaks the
R-symmetry spontaneously. Thus, gaugino masses are not prohibited.
Sfermion masses can be generated by the operator which arises from
integrating out the messengers $P$ and $\tl{P}$
\beqa \Delta K&\sim&-\(\f{g_{SM}^2}{16\pi^2}\)^2\int d^4\theta
\f{c_1}{m^2}Tr(S^\da
S)(\Phi^\da\Phi)~,\eeqa which gives  \beqa m_{\tl{f}}^2\sim&\(\f{g_{SM}^2}{16\pi^2}\)^2\f{c_1}{m^2}(F_S^\da
F_S)~. \eeqa
Gaugino masses can be generated by an
anti-instanton induced operator~\cite{yanagida2} \beqa c_2\int
d^4\theta \(\f{1}{16\pi^2}\) \f{(\Lambda_L^{\da}
)^{2N_c+1}}{m^{4N_c+2}} Tr(S^\da S)\det(\bar{D}^2 S^\da)W_aW^a \eeqa
where $\Lambda_L^{\da}$ is the holomorphic dynamical scale below the
thresholds of $P$ and $\tl{P}$. The gaugino masses \beqa
m_{gaugino}=c_2\(\f{g_{SM}^2}{16\pi^2}\)\f{(\Lambda_L^{\da}
)^{2N_c+1}}{m^{4N_c+2}} (F_S^\da F_S) (F_S^\da)^{N_Q}~,
\eeqa
are not too small because the gauge couplings are large~\cite{yanagida2}.

\section{Conclusion}
\label{sec-5}
  In this paper, we propose the SUSY $SU(7)$ unification of the $SU(3)_C\tm SU(4)_W\tm
  U(1)_{B-L}$ model. Such unification scenario has rich symmetry breaking chains
  in a five-dimensional orbifold. We study in detail the SUSY $SU(7)$ symmetry breaking
  into $SU(3)_C\tm SU(4)_W\tm U(1)_{B-L}$ by boundary conditions in a Randall-
  Sundrum background and its AdS/CFT interpretation.
  We find that successful gauge coupling unification can be achieved in our
  scenario. Gauge unification favors low left-right and unification scales with
  tree-level $\sin^2\theta_W=0.15$.
  We use the AdS/CFT dual of the conformal supersymmetry breaking scenario to
  break the remaining ${\cal N}=1$ supersymmetry. We employ AdS/CFT to reproduce the NSVZ
  formula and obtain the structure of the Seiberg duality in the strong coupling
  region for $\f{3}{2}N_c<N_F<3N_C$. We show that supersymmetry is indeed broken
  in the conformal supersymmetry breaking scenario with a vanishing singlet
  vacuum expectation value. 
\begin{acknowledgments}
We acknowledge the referee for useful suggestions. This research was supported in part by the Australian Research
Council under project DP0877916 (CB and FW), by the National Natural
Science Foundation of China under grant Nos. 10821504, 10725526 and
10635030, by the DOE grant DE-FG03-95-Er-40917, and by the
Mitchell-Heep Chair in High Energy Physics.
\end{acknowledgments}

\end{document}